\documentstyle[preprint,eqsecnum,aps]{revtex}
\tightenlines

\font\impd = cmssi12
\font\imp = cmssi10
\font\im = cmmi7

\def\u{\hbox{\impd u}}
\def\v{\hbox{\imp v}\,} 
\def\w{\hbox{\imp w}\,}
\def\x{\hbox{\imp x}\,} 
\def\y{\hbox{\imp y}\,} 
\def\z{\hbox{\imp z}\,}
\def\f{\hbox{\imp f}\,}
\def\A{\hbox{\imp A}\,} 
\def\B{\hbox{\imp B}\,} 
\def\C{\hbox{\imp C}\,}
\def\I{\hbox{\imp I}\,} 
\def\K{\hbox{\imp K}\,} 
\def\L{\hbox{\imp L}\,}
\def\N{\hbox{\im N}\,}   
\def\V{V\,}
\def\delt{{\delta}} \def\dolt{{\delta^\natural}} \def\Delt{{\delta^\sharp}}
\def\Y{{\cal Y}} \def \Z{{\cal Z}}

\begin{document} 

\title{Moving vortex in relativistic irrotational
perfect fluid or superfluid} \author{Brandon Carter, David Langlois and
Denis Priou} \address{D\'epartement d'Astrophysique Relativiste et de
Cosmologie,\\ UPR 176, Centre National de la Recherche Scientifique,\\
Observatoire de Paris, 92195 Meudon, France} \date{\today} \maketitle

\bigskip\bigskip \begin{abstract} 

Irrotational relativistic vortex configurations in uniform subsonic
motion with respect to a surrounding perfect fluid are analysed for the
purpose of application to superfluid layers in neutron stars.
Asymptotic solutions are found by asymptotically expanding the flow
equation at large distances from the vortex core and then by solving it
order by order.  The  asymptotic effective tension and energy density
that are needed for an averaged macroscopic description are thus
obtained as functions of the vortex velocity, the vortex circulation,
the asymptotic   chemical potential and of  parameters depending on the
equation of state.  \end{abstract}

\section{Introduction}

The aim of this work is to obtain large scale average values of
physically important quantities such as energy and tension for
stationary longitudinally invariant ``vortex" solutions for a
relativistic irrotational perfect fluid in an asymptotically uniform
background in a flat spacetime.  The present analysis is restricted to
the case in which only a single vortex is present. This analysis
constitutes an essential first step towards the large scale treatment
of cases in which, as discussed in the concluding section, there is an
extended  array of vortices.

The main application we have in mind is the interior layers of a
rotating neutron star, in which the bulk of the matter can be
represented to a good approximation as a relativistic superfluid at
zero temperature, whose local behaviour on a microscopic scale is that
of an irrotational perfect fluid. This means that the vorticity
resulting from the stellar rotation will be concentrated in discrete
vortex cores which may be subject to external drag forces\cite{SS95a}
that cause them to move relative to the ambient superfluid. In order to
evaluate the effect of such relative movement, the present study
generalises results of an earlier investigation \cite{cl2} that was
restricted to the axisymmetric case in which the vortex is at rest with
respect to the uniform background, so that an exact analytic
description was possible even for the inner regions of the vortex. In
that previous work, the fluid under investigation was a superfluid at
non zero temperature, which can be described by the relativistic
extension \cite{ck} of Landau's two fluid model \cite{landau}. Thanks
to the numerous symmetries of the configuration, namely invariance
under time translation, invariance under  longitudinal translation
(along the axis of the vortex) and axisymmetry, it was possible to
obtain exact analytical results for a ``cool" superfluid (Bose
condensate plus a perfect gas of phonons \cite{cl1}). Here, we still
assume invariance under time translation and longitudinal translation
but axisymmetry is lost because of the motion of the vortex with
respect to the surrounding fluid.

This loss of axisymmetry  makes it difficult to obtain exact analytic
results except in the extreme limit of a ``stiff" fluid -- in which the
speed of sound is equal to that of light -- which is relevant for
various cosmological contexts, notably the case in which the superfluid
is formed as an axionic field condensate whose vortices constitute what
are known as ``global" cosmic strings \cite{ds}\cite{gkpb}. The present
work is intended  for the rather different astrophysical context of
neutron star interiors -- in which the relevant sound speeds are
expected to be considerably (albeit not incomparably) smaller than the
speed of light. In these circumstances it is not easy to provide an
exact analytic treatment of the inner regions of a  moving vortex.
However the present work shows that it it is possible to provide a very
satisfactory description of the outer regions by an asymptotic
expansion. This description is quite sufficient for the purpose of
calculating the large scale averages (of energy, tension et cetera)
that are needed in practice for the usual astrophysical applications.

It will be convenient to work with cylindrical coordinates with respect to
which the spacetime metric $g_{\mu\nu}$ ($\rho,\sigma=0,1,2,3$) is given by
\begin{equation}
g_{\rho\sigma}dx^\rho dx^\sigma
=-c^2dt^2+d\ell^2+dr^2+r^2 d\theta^2 \ ,\label{metric}
\end{equation}
where $c$ is the speed of light, so that the
postulated stationarity and longitudinal invariance will be expressible as the
condition that the relevant physical quantities should be independent of $t$
and $\ell$. It is to be remarked that the neglect of general relativistic
curvature effects in this work is justified in so far as most neutron star
(and even cosmological) applications are concerned by the consideration that
the gravitational curvature radii involved will be extremely large compared
with the lengthscales -- of which the most important are the intervortex
separation distances --  that are relevant for the local effects considered
here. It will be supposed that the latter distances are themselves very
large compared with the lengthscales characterising the central vortex
core, whose details may consequently be neglected without significant
loss of accuracy. (Neglect of the core would not of course be justifiable
if we were concerned with cosmic strings of ``local" as opposed to 
``global" type, nor of the analogous phenomen of the 
quantised magnetic flux tubes in neutron star matter.) 

The present analysis will not allow for thermal effects, whose treatment would
require the use of the generalised Landau type two constituent superfluid
model \cite{ck}, \cite{cl2}, but whose consequences are not very
important in typical neutron star applications for which the zero temperature
limit approximation is sufficient.  Our attention will be restricted here to
superfluid models of the simplest ``barotropic" kind for which a complete set
of independent physical variables is constituted just by the components of the
4-momentum covector $\mu_\sigma$ whose dynamics is governed by the
irrotationality condition
\begin{equation}
 \mu_\sigma=\nabla_{\!\sigma\,}\varphi \ ,\label{1.2}
\end{equation}
where $\varphi$ is a gauge dependent scalar -- interpretable as a phase angle
in the underlying bosonic condensate -- with period $2\pi$ in units such that
the Dirac Plank constant $\hbar$ is set to unity. The complete system of the
equations of motion depends on the specification of the appropriate equation
of state,  giving the fluid pressure $P$ as a function of the square of the
effective mass or ``relativistic chemical potential" $\mu$ -- defined as the
magnitude of the momentum covector -- which determines the corresponding
dilatonic amplitude scalar $\Phi$ according to the prescription 
\begin{equation}
\Phi^2={2\over c^2}{dP\over d\mu^2}\ ,\hskip 1.2 cm 
c^2\mu^2=-\mu^\sigma\mu_\sigma=-\varphi_{,\sigma}\nabla^{\sigma}\varphi .
\label{1.3}
\end{equation}
This in turn determines the corresponding particle current vector $n^\sigma$,
which is given by
\begin{equation}
 n^\rho=\Phi^2\mu^\rho \ ,\label{1.4}
\end{equation}
while the associated stress energy momentum density tensor will be given by
\begin{equation}
T^{\rho\sigma}=\Phi^2 \mu^\rho\mu^\sigma+P g^{\rho\sigma} \ .\label{1.5}
\end{equation}
In conjunction with (1.2), all that is needed to complete the system of
dynamical equations of the superfluid (thereby automatically ensuring the
conservation of the stress momentum energy tensor) is the equation expressing
the conservation just of the current (1.4). This will evidently be presentable
as a non-linear wave equation in the form
\begin{equation}
\nabla_{\!\sigma}\big(\Phi^2\nabla^\sigma\varphi\big)=0 \ . \label{1.6}
\end{equation}
The special ``stiff" (Zel'dovich type) case
 is characterised by the special condition that $\Phi^2$ simply be
constant (which arises for an equation of state of the form $P\propto
\mu^2-m^2$, where $m$ is a fixed mass per particle), so that (\ref{1.6}) reduces to
a linear equation of the familiar Dalembertian form, for which the
bicharacteristic speed $c_{_{\rm S}}$ of propagation of small perturbations is
evidently equal to the speed of light $c$. For the more general class of
equations of state covered by the present treatment, the corresponding
bicharacteristic sound speed $c_{_{\rm S}}$ will be given by the formula
\begin{equation}
(c/c_{_{\rm S}})^{\, 2}=2{\mu^2\over\Phi^2}{d\Phi^2\over d\mu^2}+1\ .\
\label{1.7}
\end{equation}
For an equation of state of ``relativistic
polytropic" type $P\propto \mu^{N+1}$ where ${\N}$ is a fixed ``polytropic
index" number, one obtains the  fixed value $(c_{_{\rm S}}/c) ^{\, 2}=
1/{\N}$.

In order to determine the asymptotically contributions to the modifications
of energy and tension due to the presence of a vortex, the only information
about the uniform background state that was found to be needed in the
axisymmetric non-moving case \cite{cl2} consisted of the asymptotic
limit values, $\mu_{_\infty}^{\,2}$ and  $\Phi_{_{\infty}}^{\,2}$ say, 
of the squared chemical potential $\mu^2$ and the squared amplitude $\Phi^2$,
together with the corresponding limit value $c_{_{\infty}}^{\,2}$ of the 
squared sound speed $c_{_{\rm S}}^{\, 2}$ which it will be convenient to 
express in terms of a dimensionless non-negative parameter $\delt$ defined 
as a measure of deviation from ``stiffness" by the formula
\begin{equation}
\delt={1\over 2}\big((c/c_{_{\infty}})^2-1\big)=
\Big( {\mu^2\over\Phi^2} {d\Phi^2\over d\mu^2}\Big)
 \Big |_{_\infty}\ .\label{1.9}
\end{equation}
In the case of a simple relativistic polytrope it will be given by
$2\delt=\N-1$, which means that it will be equal to unity, $\delt=1$, in the
case $\N=3$ for which the equation of state is that of the standard model for
a gas of ultrarelativistic particles, which has been commonly used as a crude
first approximation for the treatment of neutron star matter. In the case of a
relatively moving mortex, it will be shown here that this information is not
quite sufficient, but that it is also necessary to know one more parameter,
characterising the next order of differentiation of the equation of state.
This extra parameter, $\dolt$ say, is conveniently specifiable in the form
\begin{equation}
\dolt=-{1\over 2}\biggl(\mu^2 { d (c/c_{_{\rm S}})^2)\over d\mu^2}
\biggr) \biggr|_{_\infty} =- \Biggl( \mu^2 {d\over d\mu^2}
\biggl( {\mu^2\over\Phi^2} {d\Phi^2\over d\mu^2}
\biggr)\Biggr)\Biggr |_{_\infty} \ ,\label{1.10}
\end{equation}
so as to be interpretable as a dimensionless measure of deviation from the
relativistic polytropic case. Although it is conceivable that it might be
negative, one would expect in practice that for realistic equations of state
this parameter $\dolt$ would usually be positive . 
 
In addition to this minimal information about the uniform background state,
the only other parameter needed to characterise the vortex in the
non-moving axisymmetric case is the corresponding circulation integral
$\kappa$ which (since we have set $\hbar=1$) must be a multiple of $2\pi$,
or equivalently the corresponding angular momentum $\A$ say per unit
mass, taking the latter to be the asymptotic value $\mu_{_\infty}$ of
the relativistic chemical potential $\mu$, 
provided it is taken for granted that the value of the net current
input $\I$ per unit length is zero. These conserved
quantities will be expressible as averages over a circle of radius $r$ 
(whose value can be chosen arbitrarily without affecting the result) by
the formulae
\begin{equation}
 \A={\kappa\,\over\, 2\pi\mu_{_\infty}\!}\ ,\hskip 1.2 cm
  \kappa=\oint d\varphi= \overbrace{2\pi \varphi^\prime}\, \label{1.11}
\end{equation}
and
\begin{equation}
\I=\overbrace {2\pi r\, n^\sigma\nu_\sigma}  \ ,\label{1.12}
\end{equation}
where $\,\overbrace{\,}\,$ indicates angular averaging over the chosen circle 
of radius $r$, and a prime indicates differentation with respect to the 
angle $\theta$ round the circle, while $\nu_\sigma$, with components 
(0,0,1,0) in the system (1.1), is the outgoing unit normal to the circle. 
Finally, in the case of a moving vortex, the specification of $\A$, 
and if necessary of $\I$, must evidently be supplemented by the 
specification of the velocity, $\beta$ say, of the relative motion in 
order for the characterisation of the vortex to be complete.  It is 
conceivable in principle that $\I$ might be given a non-zero value by
an artificially contrived injection process in a laboratory experiment, 
but in the natural context of neutron star vortices 
it is reasonable to assume that -- as will be postulated in the last
sections  of the present work -- the central core input
should vanish: $\I=0$. 

The main results of the present work are the calculation of the 
asymptotic form of the {\it tension},
i.e. the integrated longitudinal stress, $T$ say, and the corresponding
{\it energy per unit length} $U$ say, over a circular cross section of radius
$r$ through the vortex,
 as functions of the background parameters 
$\mu_{_\infty}^{\,2}$, $\Phi_{_\infty}^{\,2}$, $\delt$ and $\dolt$ and of the parameters
$\A$ and $\beta$ that specify the amplitude and velocity
of the vortex itself, subject to the assumption
 that the injection rate $\I$ is zero. As in the familiar non-moving case
these quantities are found to have a logarithmic radial dependence.

\section{The asymptotic expansion and its lowest order}

Our objective is to evaluate large scale averages of energy and tension for
stationary, longitudinally invariant, asymptotically uniform solutions for
an irrotational barotropic relativistic perfect fluid in flat space. 
It is not possible to solve explicitly the superfluid dynamical equation 
(\ref{1.6}) except in the special ``stiff" case where this equation 
becomes linear. Since we are ultimately interested by the asymptotic 
effect of the vortex and not the details of the flow near the core of the 
vortex, we shall try to solve this equation by using an asymptotic
expansion at large $r$. Prior to this, it is convenient to 
reexpress (\ref{1.6})  in the form
\begin{equation}
2\nabla_{\!\sigma}\nabla^\sigma\varphi+\big( (c/c_{_{\rm
S}})^{\,2}-1\big)\mu^{-2}(\nabla_{\!\sigma}\mu^2)\nabla^\sigma\varphi=0 \
.\label{1.8}
\end{equation}

At this stage, one can invoke the time translation symmetry  and the 
longitudinal symmetry, to which one can associate two Killing vectors
$k^\sigma$ and $l^\sigma$ generating respectively time and longitudinal 
space translations and corresponding to the ignorable coordinates 
$t$ and $\ell$. These two Killing vectors provide us with two 
constants of motion 
\begin{equation}
k^\sigma\mu_\sigma=-E, \quad l^\sigma\mu_\sigma={\cal L},
\end{equation}
where $E$ is interpretable as the effective energy per particle, and 
${\cal L}$ is interpretable as an effective longitudinal momentum 
per particle. Consequently, the  potential $\varphi$ has a linear
dependence on time  $t$ and longitudinal distance $\ell$ in cylindrical
coordinates.
Since there is no loss of generality in eliminating the $\ell$ dependence 
altogether, i.e.
 setting ${\cal L}=0$, by a longitudinal boost, the potential will be 
conveniently 
expressible in the form
\begin{equation}
\varphi=E\Big({\u\over c^2}-t\Big)\label{5}
\end{equation}
where  $\u$ is a function only of $r$ and $\theta$.

 In order to analyse that
asymptotic form of the flow at large $r$, we now postulate that $\u$ has an
expansion of the form
\begin{equation}
\u=\v r +\w \ln r + \x +\y {\ln r\over r}+{\z\over r}
+o\Big({1\over r}\Big) \ ,\label{6}
\end{equation}
where $\v$, $\w$, $\x$, $\y$, and $\z$ are functions of $\theta$ only. 
In order to proceed, we need the first and second partial derivatives of
$\u$ with respect to $r$ and with respect to $\theta$. To the relevant
asymptotic order, the former will be expressible as
\begin{eqnarray}
\u_r&=&\v+{\w\over r}-\y{\ln r\over r^2}+{\y-\z\over r^2}
+o\Big({1\over r^2}\Big) \ ,\label{7}\\
\u_{rr}&=&-{\w\over r^2}+2\y{\ln r\over r^3}+{2\z-3\y\over r^3}
+o\Big({1\over r^2}\Big) \ ,\label{8}
\end{eqnarray}
while the latter will be expressible as
\begin{eqnarray}
\u_\theta &=&\v ^{\prime} r +\w^{\prime} \ln r + \x^{\prime}
+\y^{\prime} {\ln r\over r}+{\z^{\prime}\over r}
+o\Big({1\over r}\Big)\ ,\label{9} \\
\u_{\theta\theta}&=&\v ^{\prime\prime} r +\w^{\prime\prime} \ln r
+ \x^{\prime\prime}+\y^{\prime\prime} {\ln r\over r}+{\z^{\prime\prime}\over r}
+o\Big({1\over r}\Big)\ ,\label{10}
\end{eqnarray}
using a prime to denote straight differentiation with respect to $\theta$.

According to (\ref{1.3}), the effective mass $\mu$ will be given by
\begin{equation}
c^2\mu^2 ={E^2\over c^4}\big(c^2-\u_r^{\ 2}-{1\over r^2}\u_\theta^{\ 2}
\big)\ .\label{11}
\end{equation}
This can be  evaluated with the help of the asymptotic forms of 
the  first partial derivatives (\ref{7}) and (\ref{9}) and we obtain
\begin{eqnarray}
{c^6\mu^2\over E^2}&=&c^2-(\v^2+\v^{\prime 2})-2\v^\prime\w^\prime{\ln r\over r}
-2{(\v\w+\v^\prime\x^\prime)\over r}-\w^{\prime 2}{(\ln r)^2\over r^2}
\nonumber \\
&&+2(\v\y-\v^\prime\y^\prime -\w^\prime\x^\prime){\ln r\over r^2}
+{2(\v\z-\v^\prime\z^\prime-\v\y)-\x^{\prime 2}-\w^2   \over r^2}
+o\Big({1\over r^2}\Big)\ . \label{14}
\end{eqnarray}
As well as the expression for $\mu^2$ itself, we shall use its partial
derivatives with respect to $r$ and $\theta$ 
for the evaluation to the required order of
both the quadratic first order differential contribution
\begin{equation}
(\mu^2)_{,\sigma}\nabla^\sigma\varphi={E\over c^2}\Big(
(\mu^2)_r\,\u_r+{1\over r^2} (\mu^2)_\theta\,\u_\theta\Big)\ ,\label{17}
\end{equation}
and the second order Laplacian contribution
\begin{equation}
\nabla_{\!\sigma}\nabla^{\sigma}\varphi={E\over c^2}\Big(\u_{rr}
+{1\over r}\u_r +{1\over r^2} \u_{\theta\theta}\Big)\ ,\label{18}
\end{equation}
in the dynamical equation (\ref{1.8}).

Independently of the particular form of the equation of state, it can be seen
that the vanishing of the coefficient of the leading term on the left of
(\ref{1.8}), namely the term of order $1/r$,  will give the requirement
\begin{equation}
\v^{\prime\prime}+\v=0\ ,\label{21}
\end{equation}
whose first integral gives
\begin{equation}
\v^2+\v^{\prime 2}=\V^2\ , \label{22}
\end{equation}
where $\V$ is a constant of integration that will be interpretable as the 
flow velocity of the uniform background. There will be no loss of generality 
in arranging for the direction of this flow to be aligned with the axis 
$\theta=0$, which means that the local solution for $\v$ will be given 
explicitly by
\begin{equation}
\v=\V\cos\theta\ .\label{23}
\end{equation}

Still independently of the particular form of the equation of state function
given by (\ref{1.7}), it can also be seen that the vanishing of the coefficient of
the next leading term on the left of (\ref{1.8}), namely the term of
order ${\ln r/ r^2}$, will give the further requirement
\begin{equation}
$$\w^\prime=0\ \label{24}
\end{equation}
as the only possibility compatible with global regularity, which simply
means that (like $\beta$) $\w$ must be a constant. We shall restrict our
attention in subsequent sections to the cases in which $\w$ is restricted to
be zero, which is the condition for there to be  no source or sink in the
vortex, i.e. for which there is no net inflow or outflow, but for the
purposes of the next section we shall provisionally continue to consider
configurations of the most general kind for which the constant $\w$ is
unrestricted.

\section{Conditions for equilibrium at intermediate order.}

In order to proceed beyond the lowest order, we must take account of the 
particular form of the function given by (\ref{1.7}), which will have an
asymptotic expansion that can be seen from (\ref{14}) to take the form
\begin{equation}
{_1\over^2}\Big( (c/c_{_{\rm S}})^{2}-1\Big)=\delt+2\dolt{\gamma^2\over
c^2}{(\v\w+\v^\prime\x^\prime)\over r}
+o\Big({1\over r}\Big)\ ,\label{25}
\end{equation}
using the usual abbreviation $\gamma^2=1/(1-\beta^2)$ with $\beta=\V/c$,
where $\delt$ and $\Delt$ are dimensionless constants defined in 
(\ref{1.9}) and (\ref{1.10}) in terms of 
asymptotic values of respectively second and third derivatives
of the equation of state function.

On substitution of the leading order dynamical conditions (\ref{22}) and 
(\ref{24}),
the expansion formula (\ref{14}) simplifies to
\begin{equation}
{c^6\mu^2\over E^2}={c^2\over\gamma^2}-2{\w\v+\x^\prime\v^\prime\over r}
-2(\y\v^\prime)^\prime{\ln r\over r^2} -{\w^2+\x^{\prime 2}+2\y\v
+2(\z\v^\prime)^\prime \over r^2}+o\Big({1\over r^2}\Big)\ .\label{27}
\end{equation}

It is now apparent from (\ref{25}) that the vanishing of the coefficient
of the leading survivimg term the left of (\ref{1.8}), namely the term of
order $1/r^2$, will give the requirement
\begin{equation}
\Big({c^2\over\gamma^2}-2\delt\v^{\prime 2}\Big)\x^{\prime\prime}+4\delt
\v\v^\prime\x^\prime +2\w\delt\big(\v^2-\v^{\prime 2}\big)=0\ .\label{32}
\end{equation}
This provides a decoupled second order differential equation for $\x$, for
which one immediately obtains a first integral expressible as
\begin{equation}
\Big({c^2\over\gamma^2}-2\delt\v^{\prime 2}\Big)\x^\prime
-2\w\delt\v\v^\prime=\A {c^2\over\gamma^3}\sqrt{1-\alpha^2} \ ,\label{33}
\end{equation}
where $\A$ is an arbitrary constant of integration which is interpretable
as proportional to the corresponding value of the total {\it circulation}
$\kappa$  round the vortex, in terms of which it is given by the relation
\begin{equation}
\A={\kappa\gamma c^2\over 2\pi E}\ ,\hskip 1.2 cm
\kappa=\oint d\varphi\ , \label{34}
\end{equation}
in which $\alpha$ is a predetermined constant given by
\begin{equation}
\alpha^2=2\beta^2\gamma^2\delt={(c/c_{_{\rm S}})^{2}-1\over 
(c/\V)^2-1}  \ ,\label{35}
\end{equation}
which, assuming the asymptotic flow velocity $\V=c\beta$ to be subsonic, will
necessarily be less than unity:
\begin{equation}
\V^2<c_{_{\rm S}}^{\,2} \qquad \Rightarrow\qquad
 \alpha^2<1\ .\label{36}
\end{equation}

Subject to this subsonicity condition, which will be taken for granted
in all that follows, the first order differential equation (\ref{33}) can in its
turn be integrated straightforewardly to give the explicit solution for $\x$ 
which is found to be simply expressible in terms of a modified angle
variable $\psi$ by 
\begin{equation}
\x=\A\gamma^{-1} \psi 
+\w\ln\sqrt{1-\alpha^2\sin^2\theta} \ ,\qquad\qquad
\psi={\rm arctan}\big\{\sqrt{ 1-\alpha^2}\,\tan\theta\big\}\label{37}
\end{equation}
The absence of a second arbitrary constant of integration from this expression
does not imply any loss of generality, because the absence of radial 
dependence in the corresponding term in (\ref{6}) means that it is only the
derivative of $\x$ but not its absolute value that is physically relevant.
For the same reason (unlike the other angular dependence functions
$\v$, $\w$, $\y$, $\z$ which must all be strictly continouous) the
function $\x$ can be discontinuous-- and indeed it necessarily will 
be so somewhere unless the circulation $\kappa$ vanishes. It can be observed 
that while the physically meaningful derivative $\x^\prime$ is a 
continuous function of $\theta$, the expression (\ref{37}) gives a value of
the potential $\x$ itself that has jump discontinuities at $\theta=\pm\pi/2$. 
(Insertion and adjustment of an additative constant of integration on the 
right of (\ref{37}) could be used to make gauge changes of $\varphi$ by which
the discontinities could be displaced elsewhere, or by which one -- 
though not both -- of them could be removed altogether.)

Again using (\ref{25}), we can go on to evaluate the coefficient of the
next leading term in (\ref{1.8}), namely the one of order $\ln r/ r^3$,
whose vanishing can be seen to be expressible as the condition
\begin{equation}
\Big({c^2\over\gamma^2}-2\delt\v^{\prime 2}\Big)\y^{\prime\prime}+8\delt
\v\v^\prime\y^\prime +\Big({c^2\over\gamma^2}+2\delt(\v^{\prime 2}-2\v^2)
\Big)\y=0\ .\label{38}
\end{equation}
We thus again obtain a decoupled second order equation, this time for
$\y$, for which it can be seen that there is a first integral given by
\begin{equation}
\Big(1-2\delt{\gamma^2\over c^2}\v^{\prime 2}\Big)\v^\prime\y^\prime+
\Big(1+2\delt{\gamma^2\over c^2}\v^{\prime 2}\Big)\v\y=\V\B\ ,\label{39}
\end{equation}
%$$\big(\gamma^{-2}-2\delt\v^{\prime 2}\big)\v\y^\prime -\big(\gamma^{-2}
%-2\delt(\v^{\prime 2}+2\v^2)\big)\v^\prime \y=\beta
%\gamma^{-2} \C \ ,\eqno(39)$$
where $\B$ is a constant of integration. The complete solution 
is hence finally obtainable in the explicit form
\begin{equation}
\y={\B\cos\theta+\C\sin\theta\over 1-\alpha^2\sin^2\theta}\ , \label{40}
\end{equation}
where $\C$ is another constant of integration. The 
regularity of this solution is guaranteed by the subsonicity condition
(\ref{36}) which ensures that the denominator does not vanish anywhere.
It is to be noted that unlike the solution (\ref{37}) for $\x$, the solution
(\ref{40}) for $\y$ is not affected by the presence of a non vanishing
source coefficient $\w$.

Although the values of the integration constants $\C$ and $\B$ appear at 
this stage to be arbitrary, it transpires that they are in fact 
predetermined by the condition of existence of a globally regular solution 
for the still unknown function $\z$. 
The latter is governed by a higher order equilibrium equation 
which we must now work out, even though it turns out that the integrals we 
shall need for evaluating the asymptotically dominant contribution to the 
energy and tension can be evaluated without any specific knowledge of the 
actual functional form of the higher order coefficient $\z$ itself.

\section{Conditions for equilibrium at the relevant higher order.}

In view of the automatic cancellation expressed by (\ref{41}), the dominant
surviving contribution to the integrals with which we are concerned will
come from the next term in (\ref{27}), which is only of order $1/r^2$.
In order to obtain the information required for
evaluating this contribution, we now consider the next leading term
in (\ref{1.8}) which (since we have already dealt with all the terms up to
$\ln r/ r^3$) is the one of order $1/ r^3$. The vanishing of the
corresponding coefficient, as evaluated using (\ref{25}), is found to be
expressible as the condition 
\begin{eqnarray}
&&\Big({c^2\over\gamma^2}-2\delt\v^{\prime 2}\Big)\z^{\prime\prime}+8\delt
\v\v^\prime\z^\prime +\Big({c^2\over\gamma^2}+2\delt(\v^{\prime 2}-2\v^2)\Big)\z
-2\Big({c^2\over\gamma^2}+\delt(\v^{\prime 2}-3\v^2)\Big)\y \nonumber \\
&&\quad -4\delt\v\v^\prime\y^\prime
-2\Big( 1+2\delt+2\dolt\gamma^2c^{-2}\v^{\prime 2}\Big)\v^\prime
\x^\prime\x^{\prime\prime}+4\Big(\delt+2\dolt{\gamma^2\over c^2}
\v^{\prime 2}\Big) \v\x^{\prime 2}\nonumber \\
&&\quad =2\v\x^{\prime\prime}\w+4\dolt{\gamma^2\over c^2}\Big(
\v\v^\prime\x^{\prime\prime}+(\v^{\prime 2}-3\v^2)\x^\prime\Big)\v^\prime\w
-4\Big(\delta+\dolt{\gamma^2\over c^2}(\v^2-\v^{\prime})\Big)\v\w^2
 \ .\label{41}
\end{eqnarray}
in which the right hand side will vanish when the source coefficient
$\w$ is set to zero.

For the sake of simplicity, let us from now on restrict our attention to the
physically important source free case, meaning that we shall impose the 
postulate
\begin{equation}
\w=0 \ . \label{42}
\end{equation}
In these circumstances the right hand side of (\ref{41}) will drop out,
leaving an equation for $\z$ that can be rewritten, with the unknown
terms grouped on the left, in the form
\begin{eqnarray}
&&\biggl(\! \Big({c^2\over\gamma^2}-2\delt\v^{\prime 2}\Big)\v\z^\prime-\Big(
{c^2\over\gamma^2}-
2\delt(\V^2+\v^2)\Big)\v^\prime\z\!\biggr)^\prime=
\biggl(\! \Big( 1+2\delt+2\dolt{\gamma^2\over c^2}\v^{\prime 2}\Big)
\v\v^\prime\x^{\prime 2}+4\delt\v^2\v^\prime\y\! \biggr)^\prime \nonumber\\
&&\qquad \qquad +\biggl(\v^2- \v^{\prime 2}-2\V^2\Big(2\delt
+\dolt{\gamma^2\over c^2} \v^{\prime 2}\Big)\!\biggr)\x^{\prime 2} +2\Big(
{c^2\over\gamma^2}-\delt(\V^2+2\v^{\prime 2})\Big)\v\y \ .\label{43}
\end{eqnarray}
The $\y$ dependent term and the term proportional to $\x^{\prime 2}$ 
at the end of the preceeding expression can be 
evaluated in the form of a derivative.
We can thus obtain a first integrated version of the equation (\ref{43}) for
$\z$ in the form
\begin{equation}
(1-\alpha^2\sin^2\theta)\cos\theta\,\z^\prime
+\big(1-\alpha^2(1+\cos^2\theta)\big)
\sin\theta\,\z={\cal F}\ ,\label{46}
\end{equation}
in terms of a known source function ${\cal F}$, from which the fully
integrated solution will be obtainable the form
\begin{equation}
\z={\f\cos\theta\over 1-\alpha^2\sin^2\theta} \ ,\label{47}
\end{equation}
where the function $\f$ is calculable by direct quadrature as a solution
of the differential relation
\begin{equation}
\f^{\,\prime}={{\cal F}\over \cos^2\theta}\ .\label{48}
\end{equation}
It can be seen that the inhomogeneous source term on the right of (\ref{46}) 
will be given by an expression of the form
\begin{equation}
{\cal F}= {\cal P}+{\cal Q}+{\cal R}\ ,\label{49}
\end{equation}
in which the first term is an automatically periodic symmetric
contribution expressible by
\begin{equation}
{\cal P}=\K +\C\Big(\ln\sqrt{1-\alpha^2\sin^2\theta}-{\cos ^2\theta
\big(1+\alpha^2\sin^2\theta\big) \over 1-\alpha^2\sin^2\theta}\Big)
 \label{50}
\end{equation}
where $\K$ is a new constant of integration, while the second term is an 
antisymmetric contribution,  with a manifestly periodic form 
expressible by
\begin{eqnarray}
{\cal Q}&=&
{\V\A^2\sin 2\theta\over c^2}\Biggl({1+\beta^2\gamma^2(2\delt^2\!-\delt+\dolt)
\over 2(1-\alpha^2\sin^2\theta) }
 -{(1-\alpha^2)\big( 1+2\delt+2\dolt\gamma^2\beta^2\sin^2\theta \big)
\over 2(1-\alpha^2\sin^2\theta)^2 }
\Biggr)\nonumber \\
&&+\B\sin 2\theta\,{1 -\alpha^2(1+\cos^2\theta)\over
2(1-\alpha^2\sin^2\theta)}.\label{51}
\end{eqnarray}
What matters most for our present purpose is the third term in (\ref{49}), which is 
an antisymmetric remainder with a not automatically periodic form given by
\begin{equation}
{\cal R}=-{\A^2\V\big(2\delt-\beta^2\gamma^2
(2\delt^2\!-\delt-\dolt)\big)\over c^2 (1-\alpha^2)^{1/2}}\psi
  +\B\sqrt{1-\alpha^2}\,\psi \ .\label{52}
\end{equation}
Unlike the other terms, the second contribution ${\cal R}$ will be subject, 
for generic parameter values, to a mismatch $\big[{\cal R}\big]_0^{2\pi}$ 
between its value at departure from the axis of motion, when  $\psi=0$,
and its value at return to this axis, when $\psi=2\pi$, since it is evident 
from (\ref{52}) that we shall have
\begin{equation}
\big[{\cal R}\big]_0^{2\pi}=2\pi\Big(
- {\A^2\V\big(2\delt-\beta^2\gamma^2
(2\delt^2-\delt-\dolt)\big)\over c^2 (1-\alpha^2)^{1/2}}  
+\B\sqrt{1-\alpha^2}\Big)  \ .\label{53}
\end{equation}

The only physically admissible solutions are those for which $\z$ and 
hence also $\cal F$ satisfies the condition of being a regular periodic
function of the angle $\theta$. Since this condition is automatically 
satisfied by the other contributions ${\cal P}$ and ${\cal Q}$ it must 
therefore also be satisfied by the third term ${\cal R}$ in (\ref{49}), which 
means that the mismatch (\ref{53}) must be made to vanish. It can thus be seen 
that the integration constant $\B$ cannot be chosen independently but that 
-- as a necessary condition for global regularity -- it must be specified
in terms of the constant $\A$ (which according to (\ref{34}) is proportional to
the circulation $\kappa$) by the relation
\begin{equation}
\B=\V\Big({\A\over c}\Big)^2{\big(2\delt-\beta^2\gamma^2
(2\delt^2-\delt-\dolt)\big)\over (1-\alpha^2)} \ .\label{54}
\end{equation}
This not only eliminates the mismatch but actually gets rid of the
remainder term altogether, so that we have
\begin{equation}
{\cal R}=0\ ,\hskip 1.2 cm {\cal F}={\cal P}+{\cal Q}\ ,\label{55}
\end{equation}
and it can also be seen to have the important secondary effect
of removing the singularity that would otherwise have arisen from the
the contribution to $\z$ via (\ref{47}) and (\ref{48}) from the antisymmetric
term ${\cal Q}$ which, by substitution of (\ref{54}), is rather
miraculously simplified so as to be given by
\begin{equation}
{{\cal Q}\over \cos^2\theta}={\A^2\V \sin 2\theta\over c^2}\,
\Biggl({\beta^2\gamma^2\big(2\delt^2+\delt+\dolt\big)\over 
(1-\alpha^2\sin^2\theta)^2}-{\alpha^2\big(2\delt-\beta^2\gamma^2(2\delt^2
-\delt-\dolt)\big)\over 2(1-\alpha^2)(1-\alpha^2\sin^2\theta) }\Biggr)
 \ .\label{56}
\end{equation}
It is evident that the corresponding contribution to the quadrature
for $\f$ as given by (\ref{48}) will thus be satisfactorily regular and periodic.
It is to be noted however that in order for $\z$ as given by (\ref{47}) to be
appropriately regular and periodic, while it is of
course necessary that $\f$ should also be periodic, it is nevertheless not
necessary that $\f$ should be regular in the strictest sense, since it
it is permissible for it to have a simple pole where $\cos\theta$ vanishes.
Such a pole will in general arise from the contribution to (\ref{47}) from
${\cal P}$, which can be seen to be expressible in the form
\begin{equation}
{{\cal P}\over\cos^2\theta}=\biggl(\tan\theta\big(\K+\C
\ln\sqrt{1-\alpha^2\sin^2\theta}\big)-\C{\psi\over\sqrt{1-\alpha^2} 
}\biggr)^\prime\ . \label{57}
\end{equation}
In addition to the admissible pole term proportional to $\tan\theta$,
the differentiand on the right of (\ref{57}) also contains an inadmissibbly
non-periodic term proportional to the modified angle coordinate $\psi$
whenever the coefficient $\C$ is non zero. It can thus be seen that
as well as fixing the constant of integration $\B$ by the condition
(\ref{54}) the requirement of global regularity also fixes the other constant
of integration introduced in (\ref{40}) by the simpler condition
\begin{equation}
\C=0\ .\label{58}
\end{equation}

The solution for $\z$ that is thus finally obtained from (\ref{47})
 and (\ref{48})  
will be given explicitly by
\begin{eqnarray}
\z&=&{\V\A^2 \cos\theta\over c^2}\,
\Biggl({\beta^2\gamma^2\big(2\delt^2+\delt+\dolt\big)\over 
\alpha^2(1-\alpha^2\sin^2\theta)^2}+{\big(2\delt-\beta^2\gamma^2(2\delt^2
-\delt-\dolt)\big)\ln\sqrt{1-\alpha^2\sin^2\theta}
\over (1-\alpha^2)(1-\alpha^2\sin^2\theta) }\Biggr)\nonumber \\
&& +{\K\sin\theta +\L\cos\theta\over 1-\alpha^2\sin^2\theta} \,
\ , \hskip 2 cm\ \label{59}
\end{eqnarray}
where $\L$ is yet another constant of integration, which, like $\K$,
presumably depends on the nature of the inner core. However what
matters for the purpose of the following sections is not the particular 
functional form of the higher order angular coefficient $\z$, but just 
the conditions (\ref{54}) and (\ref{58}) that are necessary and sufficient for its 
global regularity, and that specify the specific form of the 
intermediate order angular coefficient $\y$ which is what we really 
need to know.

\section{Asymptotic averages.}
 
Our purpose is to obtain asymptotic averages of quantities $Q$ that
will of the kind discussed in the appendix of the preceeding 
work \cite{cl2} in the sense that -- as in the prototype case of the
pressure $P$ -- they can be postulated to have a smooth functional 
dependence on the chemical potential, and will thus have a deviation 
from the corresponding uniform asymptotic limit value $Q_\infty$ 
that will have a form given by
\begin{equation}
 Q-Q_{_\infty}={dQ\over d\mu^2}\Big|_\infty(\mu^2-\mu_{_\infty}^{\,2})
+{1\over 2}{d^2 Q\over (d\mu^2)^2}\Big|_{_\infty}
(\mu^2-\mu_{_\infty}^{\,2})^2+o\Big((\mu^2-\mu_{_\infty}^{\,2})^2\Big)
 \ .\label{60}
\end{equation}
The relevant deviation of the square of the chemical
potential (or effective mass) variable $\mu$ will itself be given,
according to (\ref{27}), by
\begin{equation}
{c^2(\mu^2-\mu_{_\infty}^{\,2})\over\gamma^2\mu_{_\infty}^{\,2}}=
-2{\v^\prime\x^\prime\over r}-2(\v^\prime \y)^\prime
{\ln r\over r^2}-{\x^{\prime 2}+2\v\y+2(\v^\prime\z)^\prime\over r^2}
+o\Big({1\over r^2}\Big)\ ,\label{61}
\end{equation}
This differs substantially from the corresponding expression in our
previous analysis \cite{cl2} in which both $\v$ and $\y$ vanished
(thereby also incidentally eliminating the term involving $\z$) so that 
the leading terms of order $1/r$ and of order $\ln r/ r^2$ were absent,
leaving just the term of order $1/r^2$ with coefficient simply given
by $\x^{\prime 2}$. Its effect is to introduce new leading terms
of order $1/r$ and of order $\ln r/ r^2$ in the resulting expression
for the asymptotic deviation of the quantity $Q$, which will be given by
\begin{equation}
Q-Q_{_\infty}=-{\gamma^2\over c^2}\mu_{_\infty}^{\,2}
{dQ\over d\mu^2}\Big|_\infty \biggl(2{\v^\prime\x^\prime\over r}
+2(\v^\prime \y)^\prime {\ln r\over r^2}
+{\x^{\prime 2}+2\v\y+2(\v^\prime\z)^\prime\over r^2}\biggr)
+2{\gamma^4\over c^4} \mu_{_\infty}^{\,4} 
{d^2 Q\over (d\mu^2)^2}\Big|_{_\infty}{\big(\v^\prime\x^\prime\big)^2
\over r^2}+o\Big({1\over r^2}\Big)\ ,\label{62}
\end{equation}

The long range leading order terms in (\ref{62}) will however cancel out, leaving 
a non zero contribution only from the short range term of order $1/r^2$ when 
we take the relevant average $\overline Q$, as evaluated over a circular 
section of radius $r$ through the vortex. Such an average will be given 
by an expression of the form
\begin{equation}
\overline Q={2\over r ^2}\int^r\overbrace Q r dr
={r_{_\odot}^{\, 2}\over r^2}\,\overline Q_{_\odot}+{2\over r ^2}
\int^r_{r_{_\odot}}\overbrace Q r dr \,\label{63}
\end{equation}
where $\overline Q_{_\odot}$ is the average over an inner core of radius
$r_{_\odot}$ and
where, for any given radius $r$, the quantity $\overbrace Q$ is 
the corresponding angular average, as given by
\begin{equation}
\overbrace Q={1\over 2\pi}\oint Q\, d\theta\ .\label{64}
\end{equation}
Since the coefficient of term of order $1/r$ is antisymmetric with
respect to $\theta$ its angular average will evidently vanish. The angular
averages of the term of order $\ln r/r^2$ and of the term involving
$\z$ will also vanish because they have the form of derivatives of
periodic fuctions. The angular average of the deviation in which we
are interested will thus be given just by
\begin{equation}
\overbrace{Q-Q_{_\infty}}=-\Biggl(\mu_{_\infty}^{\,2}{dQ\over d\mu^2}
\Big|_{_\infty}{\gamma^2\over c^2} \overbrace{ (\x^{\prime 2}+2\v\y) }
-2 \mu_{_\infty}^{\,4}{d^2 Q\over (d\mu^2)^2}\Big|_{_\infty}
{\gamma^4\over c^4}\overbrace{\v^{\prime 2}\x^{\prime 2}}\Biggr){1\over r^2}
+o\Big({1\over r^2}\Big)\ .\label{65}
\end{equation}

Since the preceeding outcome is only of order $1/r^2$, it can be seen
that the required average will have an asymptotic behaviour of exactly the 
same kind as was obtained in the previous analysis \cite{cl2} of the 
non-moving case, i.e. it will be described by an expression of the form
\begin{equation}
\overline{Q}-Q_{_\infty} = {\widetilde Q\,  \A^2\over c^2}
\,  {\ln r\over r^2} +o\Big({\ln r\over r^2}\Big)  \ ,\label{66}
\end{equation}
with a constant coefficient $\widetilde Q$ which will have the same 
dimensionality as the quantity $Q$ itself, where $\A$ is a constant 
normalisation factor, interpretable as the {\it average angular momentum 
per unit mass}, that is specified independently of $Q$ by the formula
\begin{equation}
 \A={\kappa\,\over\, 2\pi\mu_{_\infty}}\ ,\label{67}
\end{equation}
which means that it is identifiable with the integration constant introduced
in (\ref{34}). The explicit formula for the asymptotic deviation coefficient
$\widetilde Q$ that is defined in this way will however be significantly 
more complicated in general than in the non-moving case, for which it 
was found to be given simply by
$\widetilde Q=-2\big(\mu^2 dQ/d\mu^2\big)|_{_\infty}$ \cite{cl2}.
The generalisation that is needed for the case of a moving vortex, i.e.
one with a non-zero value of $\beta$, can be read out from (62) as
\begin{equation}
\widetilde Q=-2\Y\mu_{_\infty}^{\,2}{dQ\over d\mu^2}\Big|_\infty
 +4\Z \mu_{_\infty}^{\,4}{d^2 Q\over (d\mu^2)^2}\Big|_{_\infty}
 .\label{68}
\end{equation}
with dimensionless coefficients $\Y$ and $\Z$ given by
\begin{equation}
\Y=
{\gamma^2\over \A^2}\big( \overbrace{ \x^{\prime 2}}+\overbrace{2\v\y}
\big) \hskip 1.2 cm \Z=
{\gamma^4\over c^2\A^2}\overbrace{\v^{\prime 2}\x^{\prime 2}}\ .\label{69}
\end{equation}
To evaluate these coefficients, we use the explicit expressions
obtained from (\ref{33}) and  from (\ref{40}) and (\ref{54}). We get 
\begin{equation}
\Z ={\beta^2\gamma^2\over 2(1-\alpha^2)^{1/2} }\ ,
\hskip 6 cm \label{73}
\end{equation}
and
\begin{equation}
\Y={2-\alpha^2\over 2 (1-\alpha^2)^{1/2}} + {(1-\sqrt{1-\alpha^2})
\big(\alpha^2-\alpha^4+\gamma^2\alpha^2+2\beta^4\gamma^4\dolt\big)\ 
\over \alpha^2(1-\alpha^2)} \ .\label{74}
\end{equation}
The latter expression is rather unweildy even in the ``polytropic" case 
characterised by $\dolt=0$, but in the stiff limit characterised by 
$\delt=\dolt=0$ it simply gives $\Y=1$, independently of the velocity 
$\beta$. 

The applications with which we are principally concerned are the
pressure $P$ for which (\ref{68}) gives
\begin{equation}
\widetilde P=-c^2\mu_{_\infty}^{\,2}\Phi_{_\infty}^{\,2}
\big(\Y-2\Z\delt\big)\ ,\label{75}
\end{equation}
and the squared amplitude $\Phi^2$ itself for which (\ref{62}) gives
\begin{equation}
\widetilde{\Phi^2}=-2\Phi_{_\infty}^{\,2}\Big(\Y\delt
-2\Z(\delt^2-\delt-\dolt)\Big) \ .\label{76}
\end{equation}
These quantities are all that is needed to evaluate the average deviations
of the longitudinal components  
\begin{equation}
n_i=\Phi^2\mu_i\label{77}
\end{equation}
of the conserved current and
\begin{equation}
T_{ij}=\Phi^2\mu_i\mu_j+Pg_{ij}\label{78}
\end{equation}
of the stress energy momentum tensor,
where the Latin indices run over the values 0,1 in the coordinate
system (\ref{metric}), with respect to which the gradient components 
\begin{equation}
\mu_i=\varphi_{,i}\label{79}
\end{equation}
take the constant values
\begin{equation}
\mu_{_0}=-E  \ , \hskip 1.2 cm \mu_{_1}=0\ ,\label{80}
\end{equation}
so that we simply obtain
\begin{equation}
\widetilde n_i=\widetilde{\Phi^2}\mu_i\label{81}
\end{equation}
and
\begin{equation}
\widetilde{T}_{ij}=\widetilde{\Phi ^2}\mu_i \mu_j +
\widetilde{P} g_{ij}.
\label{82}
\end{equation}

The general expression (\ref{66}) is not limited to scalar functions 
of $\mu^2$ but can be also applied to other quantities. This will however
require more work because the intermediate formulas such as (\ref{65}) are
no longer applicable. Interesting examples of this kind are the 
transverse current components and the transverse-transverse components of the
stress-momentum tensor. Although the resulting asymptotic coefficients
are mathematically well defined,  the physical interpretation of these
quantities is less obvious than the components treated above so that 
we have preferred to relegate these results in an appendix.

\section{Conclusions}
The present work has given asymptotic solutions for a relativistic vortex 
moving with respect to  a surrounding irrotational perfect fluid (or, 
equivalently, a zero temperature superfluid). We have obtained 
the effective energy density, which can be expressed in the form 
\begin{equation}
T_{eff}(r)={\kappa^2\over 4 \pi^2}\Phi_\infty^2\left(\Y- 2\delt\Z\right)
 {\ln r\over r^2} +o\Big({\ln r\over r^2}\Big)
\end{equation}
and the effective tension, which can be given by
\begin{equation}
U_{eff}(r)={\kappa^2\over 4 \pi^2}\Phi_\infty^2\left[\left(1-2\gamma^2\delt\right)
\Y+\left(4\gamma^2(\delt^2-\delt-\dolt)-2\delt\right)\Z\right]
 {\ln r\over r^2} +o\Big({\ln r\over r^2}\Big),
\end{equation}
where $\Y$ and $\Z$ are given by (\ref{73}) and (\ref{74}).

The present work constitutes an intermediate step that is needed for
the construction of a realistic modelization of the relativistic
dynamics of a neutron star interior. Like a superfluid in a rotating
bucket, the neutron superfluid in a rotating neutron star minimizes its
free energy by forming an array of quantized vortices that, on a
scale larger than the intervortex separation length, simulates a rigid body
rotation. Because of the huge number of vortices in a neutron star, 
it is not practicable to keep track of all individual vortices, so a
macroscopic description requires the use of a model involving
a vorticity 2-form
\begin{equation}
 w_{\rho\sigma}=2\nabla_{[\rho}\mu_{\sigma]}, 
\end{equation} 
that is not supposed to be the small scale local vorticity field,
(which vanishes outside the microscopic vortex cores) but that is to be
interpreted as the large scale average over a neighbourhood extending
across a many vortices.

The previous analysis \cite{cl2} of a relativistic vortex configuration in
a background without relative motion provided the basis for the
specification of a particularly simple model\cite{cl3} within the general
category that is needed for such a description.  Such models are
derivable from a Lagrangian type master function
$\Lambda\{n^\rho,w_{\rho\sigma}\}$ that depends on the three scalar
quantities that can be built from the (macroscopically averaged) current
density $n^\rho$ and the (macroscopically averaged) vorticity
$w_{\rho\sigma}$, namely the magnitude $n$ of the particle current
vector $n^\rho$ itself, the scalar magnitude $w$ of the vorticity
covector $w_{\rho\sigma}$ and the magnitude $\zeta$ of the associated
Joukowsky lift force density vector $\zeta_\rho$ as defined by
\begin{equation}
n^2=-n^\rho n_\rho\ , \hskip 1 cm w^2={1\over 2}
w_{\rho\sigma}w^{\rho\sigma}\ , \hskip 1 cm
\zeta^2=\zeta_\rho\zeta^\rho \ ,
\end{equation}
 where the Joukowsky vector is
defined as
\begin{equation}
\zeta_\rho=w_{\rho\sigma}n^\sigma\ .
 \end{equation}
This vector is interpretable as representing the volume density of
force that would be exerted on the vortices as an expression of the
Magnus effect, by the relative flux  of the fluid according to the
simple formula originally derived by Joukowsky  for flow past a long
aerofoil. The earlier analysis \cite{cl2} was sufficient to unambiguously
determine the appropriate form for the equation of state for the
function $\Lambda\{n,w,\zeta\}$ only in the limit for which $\zeta$
vanishes.

As convenient ansatz for use as a provisional approximation, it has been
suggested\cite{cl3} that this special limit form should be extrapolated by
assuming that there is no dependence on $\zeta$ at all, even when
$\zeta$ is non-zero, a supposition that is mathematically justifiable
for a fluid obeying the ``stiff'' Zel'dovich type equation of
state, but not in general.  To obtain a more accurate treatment for an
arbitrarily compressible fluid it will ultimately be necessary to use a
more sophisticated ansatz taking account of the effect of the relative
flow that will be present when $\zeta$ is non-zero. The present
analysis of the effect of relative flow on an individual vortex is an
indispensible firststep towards the achievement of what is required.
What still remains to be done is to extrapolate the present analysis
to the case in which there is not just one vortex but an extended array
of parallely aligned vortices. In the absence of relative flow at large
distance, i.e. in the case corresponding to vanishing $\zeta$, such an
extrapolation was almost trivial, at least in the large separation
limit that is relevant, because of the axial symmetry of the individual
vortex solution.  However  for the non axisymmetric configurations
considered in the present work the required extrapolation will not be
so straightforward.

It is to be remarked that the effect analysed in the present work is
not the only kind that needs to be examined in greater detail in order
to improve the accuracy of the recently proposed model 
\cite{cl3} for the
treatment of relative flow past the vortices in neutron stars. Whereas
the modification of the stress - momentum - energy tensor considered
here is likely to be very small, a potentially more important
correction arises from the friction between the vortices and the
``normal'' matter that will be present, not just due to thermal effects
(which in typical neutron stars will be very small) but also, even in the
zero temperature limit, due to the fact that the matter will not be
entirely constituted just by neutrons, but will also include a small but
significant fraction in the form of protons, together with a gas of
charge neutralising electrons whose scattering from the vortices can
provide a dynamically important friction drag\cite{SS95a}.

\section{Appendix: Asymptotic averages of transverse current components.}

In this appendix, we compute the asymptotic formula (\ref{66}) for quantities
which are not  scalar functions of $\mu^2$ and therefore  for which the 
method used in the main body does not apply.
 The first important example is that of the tranverse components 
with respect to Minkowskian coordinates characterised by
\begin{equation}
 x^{_0}=t\ , \ \ x^{_1} =\ell\ ,\ \ x^{_2}=r \cos\theta\ ,\ \
x^{_3} =r\sin\theta \label{83}
\end{equation}
of the conserved current $n^\sigma$. Since these components have no 
dependence on the longitudinal coordinates $x^i$ ($i= 0,1$) the
current conservation law automatically reduces from four to two dimensional 
form,
\begin{equation}
\nabla_{\!\sigma} n^\sigma=0 \hskip 1 cm \Rightarrow \hskip 1 cm
 \nabla_{\! a} n^a=0 \ , \label{84}
\end{equation}
so that we obtain
\begin{equation}
\nabla_{\! b} (x^a n^b)=n^a \ ,\label{85}
\end{equation}
using early Latin letters for the transverse indices, $a,b=2,3$.
For any 2-dimensional transverse section $\Sigma$ bounded by a curve
$s$ with outgoing unit normal having components $\nu_a$, Green's theorem
thus provides us with the relation
\begin{equation}
\int_\Sigma n^a d\Sigma=\oint_s x^a n^b \nu_b\, ds \ .\label{86}
\end{equation}
In the case of a circular section with boundary characterised by
\begin{equation}
x^a=r\nu^a\ ,\hskip 1 cm \nu^{_2}=\cos \theta\ ,\hskip 1 cm
\nu^{_3}=\sin\theta \label{87}
\end{equation}
so that $ds=r d\theta$, the preceeding formula can be used to see that 
the corresponding average -- defined as in (\ref{63}) -- will be expressible
just in terms of an angular average -- as defined in (\ref{64}) -- over the 
section boundary in the form
\begin{equation}
\overline{ {n^{\,} }^{\! a} }=2\overbrace{\nu^a \nu_b\, n^b }\ . \label{88}
\end{equation}
In the same way, for the stress energy momentum tensor we obtain
\begin{equation}
\nabla_{\!\sigma} T^\sigma_{\,\ \rho}=0 \hskip 1 cm \Rightarrow \hskip 1 cm
 \nabla_{\! a}  T^a_{\,\ \rho}=0 \ , \label{89}
\end{equation}
and hence
\begin{equation}
\overline{ {T^{\,} }^{\! a}_{\, \rho} } =2\overbrace{ \nu^a\nu_b\,
T^b_{\,\ \rho} }\ .\label{90}
\end{equation}

Using
\begin{equation}
n_a=\Phi^2\mu_a\ , \hskip 1.2 cm \mu_a\nu^a={E\over c^2}\u_r\label{91}
\end{equation}
we obtain
\begin{equation}
n^a\nu_a=\gamma\mu_{_\infty}\Phi_{_\infty}^{\,2}\Big(\v-2\delta
{\gamma^2\over c^2}
{\x^\prime\v\v^\prime\over r}-\big(\y+2\delta{\gamma^2\over c^2}
(\y\v^\prime)^\prime\v\big){\ln r\over r^2}\Big)+O\Big({1\over r^2}\Big)\ .
\label{92}
\end{equation}
Since it is evident from symmetry considerations that the angular average
$\overbrace{\x^\prime\v\v^\prime\nu^a}$ must vanish, we are left with
a relation of the standard form (\ref{66}), namely
\begin{equation}
\overline { {n^{\,} }^{\! a} }  -n^a_{_\infty} =
 \widetilde{n^a}\,  {\A^2\over c^2} \,  {\ln r\over r^2}
+o\Big({\ln r\over r^2}\Big)  \ ,\label{93}
\end{equation}
with
\begin{equation}
\widetilde{n^a}= -n_{_\infty}{\gamma c^2\over\A^2}\, 2\overbrace{\big(
\y+2\delta{\gamma^2\over c^2}(\y\v^\prime)^\prime\v\big)\nu^a}\ ,
\label{94}
\end{equation}
in which the components of the relevant angular average are calculable
from (\ref{23}) and (\ref{40}). 
Using the formula (\ref{54}) for $\B$, and noting that $\C$ vanishes
by (\ref{58}), one finally obtains
\begin{equation}
\widetilde{n^x}={2\big(2\delt-\beta^2\gamma^2
(2\delt^2-\delt-\dolt)\big)\over (1-\alpha^2)}
\biggl(1- {1-\sqrt{1-\alpha^2}\over\alpha^2}\biggr)n_{_\infty}^{\, x},
\quad \widetilde{n^y}=0
\ .\label{96}
\end{equation}

Having arrived at this result, it behoves us to remark however that the 
physical significance of the transverse current component average 
$\overline{ n^a}$ thus obtained is rather less obvious than that of the 
corresponding longitudinal average $\overline{ n^i}$ obtained in the 
previous section, which is directly interpretable as being proportional to 
the current flux through the circular section across the cylindrical
vortex cell under consideration. The corresponding transverse parts of the 
current flux associated with cylindrical vortex cells in a honeycomb lattice 
are also of obvious physical interest, but they are not directly definable 
as averages of the kind considered above, so their evaluation will require 
a rather different kind of calculation that will be left for a future 
article.

The same question of physical significance arises for the averages of the 
transverse components of the stress energy momentum tensor, for which an
analysis of the same kind as that given in the preceeding paragraph also 
leads to an equation of the standard form, namely
\begin{equation}
\overline { {T^{\,} }^{\! a}_{\,\rho} }  -T^{\, a}_{\!_\infty \rho} =
\widetilde{ {T^{\,}}^{\! a}_{\ \rho}}\,  {\A^2\over c^2}\, {\ln r\over r^2}
+o\Big({\ln r\over r^2}\Big)  \ .\label{97}
\end{equation}
In the particular case of the the cross components $T^{ai}$ this is 
obvious because they are expressible in the form 
\begin{equation}
T^{ai}=\mu^a n^i=n^a\mu^i\label{98}
\end{equation}
in which the longitudinal momentum components $\mu^i$ are constant. It is 
thus evident that the corresponding averages -- which have a direct physical
interpretation in terms of longitudinal fluxes of transverse momentum -- will simply 
be proportional to those of the current, being given by
\begin{equation}
\widetilde{T^{ai}}=-{2\big(2\delt-\beta^2\gamma^2
(2\delt^2-\delt-\dolt)\big)\over (1-\alpha^2)}
\biggl(1- {1-\sqrt{1-\alpha^2}\over\alpha^2}\biggr) n_{_\infty}^{\, a}
\mu^i \ .\label{99}
\end{equation}
For the derivation of the corresponding formulae for the purely trans verse
components $T^a_{\,\ b}$ -- which do not have such a straightforward 
physical interpretation -- a little more work is required since (\ref {86})
no longer applies. One must then consider the product of the 
asymptotic expansion of $\Phi^2$, given by (\ref{63}), with that of the 
$\mu_a$, given by (\ref{7}) and (\ref{9}), in order to obtain an  
asymptotic expansion at the required order for the  part 
$\Phi^2\mu_a\mu_b$ of the components $T_{ab}$, the pressure part being 
already known via (\ref{75}). One obtains finally
\begin{equation}
\widetilde{T_{xx}}=c^2\mu_{_\infty}^{\,2}\Phi_{_\infty}^{\,2}\left[
{\alpha^4-2\alpha^2-2\beta^4\gamma^4(\delt+\dolt)\over \sqrt{1-\alpha^2}}
+(1-\sqrt{1-\alpha^2})\big(2+\beta^2\gamma^2-\alpha^2-2(\beta^4\gamma^4\dolt
/\alpha^2)\big)\right],
\end{equation}
\begin{equation}
\widetilde{T_{xy}}=0,
\end{equation}
\begin{equation}
\widetilde{T_{yy}}=-c^2\mu_{_\infty}^{\,2}\Phi_{_\infty}^{\,2}
{(1-\sqrt{1-\alpha^2})
\big(\alpha^2-\alpha^4+\gamma^2\alpha^2+2\beta^4\gamma^4\dolt\big)\ 
\over \alpha^2(1-\alpha^2)}.
\end{equation}
In the no-motion limit $\beta=0$ or in the stiff fluid limit 
$(\delt=\dolt=0)$, 
the coefficients $\widetilde{T_{xx}}$ and $\widetilde{T_{yy}}$ vanish 
as one could have expected from the previous work \cite{cl2}.

\end{document}